\documentclass{article}

\usepackage{amsfonts, amsmath, amssymb, amsbsy, graphicx}
\usepackage[utf8]{inputenc}
\usepackage[english, german]{babel}
\usepackage[T1]{fontenc}

\def\degr{\hbox{$^\circ$}}

\begin{document}
\selectlanguage{english}

\begin{center}
\textbf{\LARGE Sonneberg Sky Patrol Archive –- Photometric Analysis}
\end{center}

\begin{center}
\textbf{Milan Spasovic, Christian Dersch, Christian Lange, Dragan Jovanovic, Andreas Schrimpf}
\end{center}

\begin{center}
{\it
\noindent History of Astronomy and Observational Astronomy, Physics Department, \\Philipps--Universität Marburg }
\end{center}

\begin{abstract}
The Sonneberg Sky Patrol archive so far has not yet been analyzed systematically. In this paper we present first steps towards an automated photometric analysis aiming at the search for variable stars and transient phenomena like novae. Early works on the sky patrol plates showed that photometric accuracy can be enhanced with fitting algorithms. The procedure used was a manually supported click-and-fit-routine, not suitable for automatic analysis of vast amount of photographic plates. We will present our progress on deconvolution of overlapping sources on the plates and compare photometric analysis using different methods. Our goal is to get light curves of sufficient quality from sky patrol plates, which can be classified with machine learning algorithms. The development of an automated scheme for finding transient events is in progress and the first results are very promising.
\end{abstract}

\section{Introduction}

The Sonneberg Plate archive with more than 275,000 plates is one of the largest plate collections in the world. Photometric observations in Sonneberg took place from about 1925 to 1995 and can be divided into two programs, Field Patrol and Sky Patrol. The Sky Patrol campaign was launched to provide a continuous record of the northern hemisphere in two different colors, some parts are covered with more than 3,000 photographic plates over a time period of 50 years. Thus, the Sky Patrol archive is very valuable source for studying the variable sky, especially of variations on long time scales.

Although more than 7 million photographic plates exist taken in the 20th century all over the world, efforts on automated analysis started just in the last decade. The most prominent pipelines for automated analysis, sorted alphabetically, are DASCH \cite{dasch}, PyPlate \cite{pyplate} and VaST \cite{vast}. These pipelines are based on Source Extractor (SExtractor), a package developed for CCD photometry, to determine star intensities or magnitudes from photographic plates.

Kroll et al. \cite{Kroll93, Vogt04} had chosen another approach. They fitted a model with a logarithmic intensity characteristics to single isolated stars on digitized Sonneberg photographic plates. For selected stars they were able to determine the brightness with an accuracy between 0.07 and 0.12 mag. However, the analyzed stars were manually selected and crowded fields could not be fitted.

In the following we will discuss two different methods we used for photometry of the Sonneberg Sky Patrol plates. Before analyzing the plates, astrometric solutions were calculated with solve-field \cite{solve} on a small sections (1\degr $\times$ 1\degr\ or 2\degr $\times$ 2\degr) of the images.

\section{Photometry using Source Extractor}

The SExtractor package supplies coordinates, magnitudes, ellipticity etc.\ for every resolved object found in an astronomical image. For analyzing digitized photographic plates SExtractor can be switched to the PHOTO mode \cite{sextractor}. In this mode data are preprocessed in order to correct the logarithmic characteristics of photographic plates
\begin{equation}
I = \frac{\gamma}{\ln 10} 10^{-0.4 \cdot m_{0}} 10^{\frac{D}{\gamma}} 
\label{exponential}
\end{equation}
\noindent where $ D $ is the pixel value, $ m_{0} $ the zero point of magnitudes and $ \gamma $ the contrast index of the emulsion. The photometric procedures of SExtractor remain unchanged after transformation of the data.

One can also perform PSF photometry with SExtractor. Therefore one uses PSF Extractor (PSFEx) to create a model of a point spread function for star images. However, this fails on photographic plates due to the non linear characteristics, which produces star images of varying widths in dependence of the magnitudes. Unfortunately PSFex can not transform the data before calculating the point spread function as in SExtractor. Therefore we applied the exponential transformation of the digitized images (eqn. \ref{exponential}) as a first step and then used PSFex and SExtractor for PSF photometry.

\begin{figure*}[htp!]
	\centering
	\hspace*{\fill}
	\includegraphics[width=0.35\textwidth]{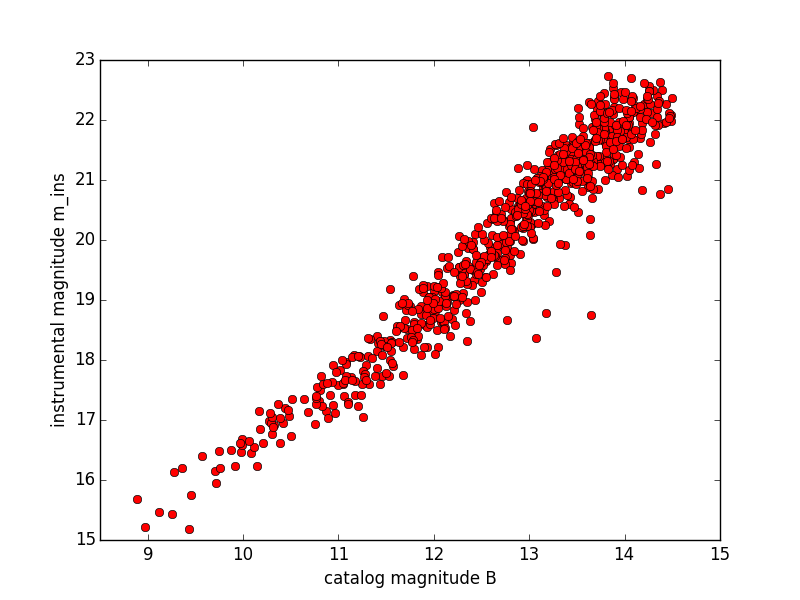}
	\hfill
	\includegraphics[width=0.35\textwidth]{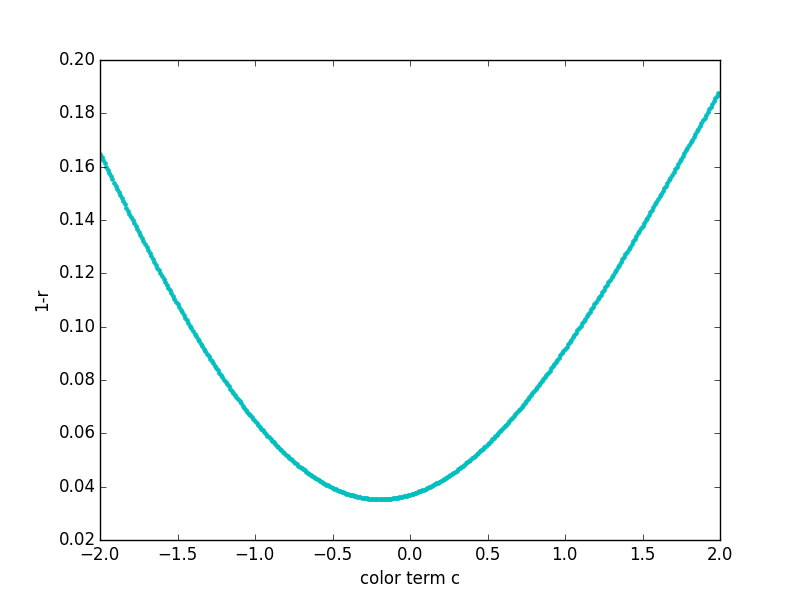}
	\hspace*{\fill}
	\caption{Calibration and color term. Left: dependency between catalog magnitude $B$ and instrumental magnitude from SExtractror. Right: $1 - r$ in dependence of color term $c$. }
	\label{calib}
\end{figure*}

For magnitude calibration the APASS (AAVSO Photometric All-Sky Survey) catalog  \cite{APASS} was chosen as reference. In the left part of fig. \ref{calib} the instrumental magnitudes are plotted in dependence of the catalog magnitudes. We analyzed the blue sensitive plates. However, their spectral sensitivity does not match the one of the Johnson B filters used by APASS. Thus, the spectral sensitivity has to be corrected by a color term as proposed by DASCH \cite{dasch}

\begin{equation}
m_{\mathrm{cat}} = B_{\mathrm{cat}} + c(B_{\mathrm{cat}} - V_{\mathrm{cat}})
\label{dscheqn}
\end{equation}

\noindent where $m_{\mathrm{cat}} $ is APASS catalog magnitude transformed in plate system, $ B_{\mathrm{cat}} $ and $ V_{\mathrm{cat}} $ APASS Johnson $B$ and $ V $ magnitude respectively and $ c $ is the color term.

Since the linearity of the characteristics nearly has been restored by the applied transformation the best correlation between instrumental and catalog magnitude could be determined by 
optimizing the Pearson correlation coefficient $r$ while varying the color term (right part of fig.\ \ref{calib}). This method is faster than repeatedly fitting the calibration curve for different values of color terms and analyzing the scattering of the residuals. And, the Pearson coefficient curves are much smoother, which enables a reliable determination of the minima.

\begin{figure*}[htp!]
	\centering
	\includegraphics[width=0.70\textwidth]{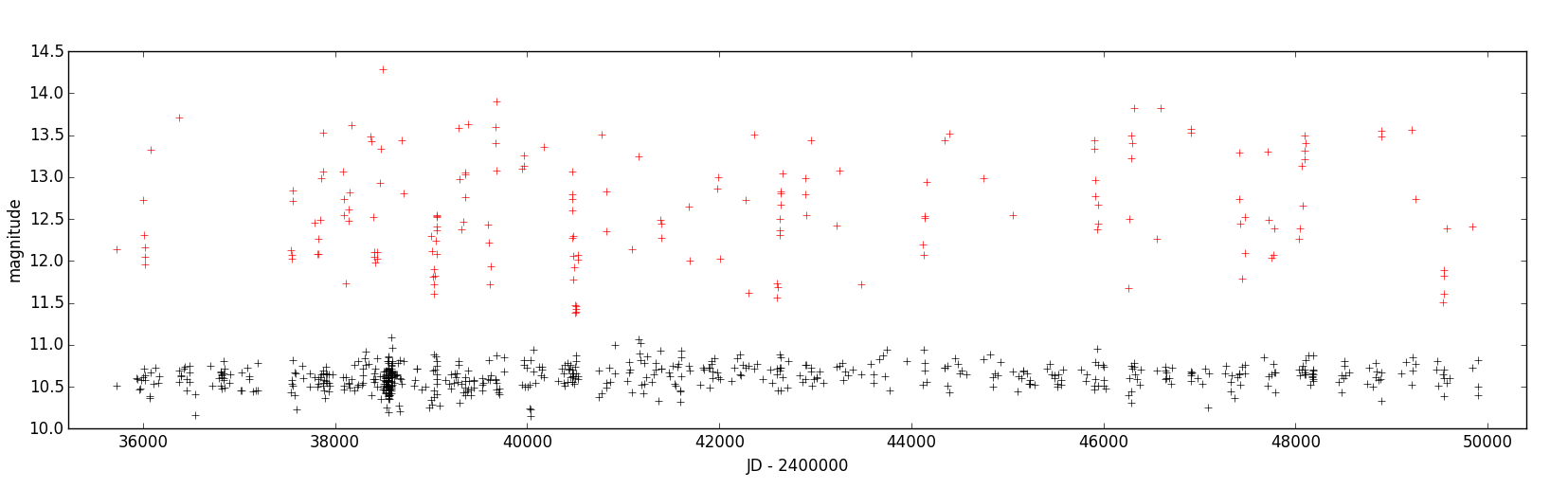}
	\includegraphics[width=0.29\textwidth]{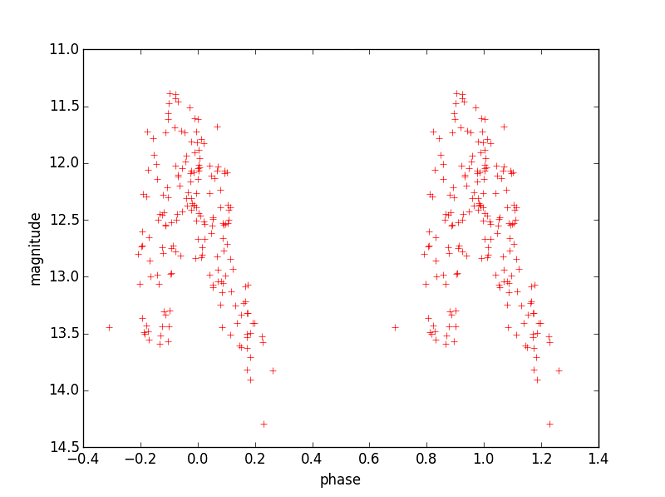}
	\caption{Lightcurves. Left: lightcurve of a constant star, rms of 0.14 (black) and lightcurve of a variable star, rms of 0.6 (red). Right: ST Lyr, folded lightcurve of variable at the known period of 300.5 days. }
	\label{lightcurve}
\end{figure*}

Photometric calibration was calculated for every single plate separately which allowed us to generate calibrated light curves for given objects. Photometric rms values of ~0.15 mag were achieved for resolved objects that show no variability (fig. \ref{lightcurve}). A larger rms value usually is evidence of a variable star, e.g. ST Lyr in fig.\ \ref{lightcurve}.

\section{Adapted PSF Fitting}

We also made some effort in fitting the digitized Sky Patrol plates with astropy \cite{astropy} using the model

\begin{equation}
f(x,y) \sim \log \left( A \cdot e^{ - \left[ \frac{{(x-a)}^{2}}{\sigma_{x}^{2}} -2 \rho \frac{(x-a)(y-b)}{\sigma_{x} \sigma_{y}} - \frac{{(y-b)}^{2}}{\sigma_{y}^{2}} \right]} + A_{0} \right)
\end{equation}
\noindent
where $ f(x,y) $ is a pixel value of digitized image.  

\begin{figure*}[htp!]
	\centering
	\scalebox{1}[0.85]{\includegraphics[width=0.32\textwidth]{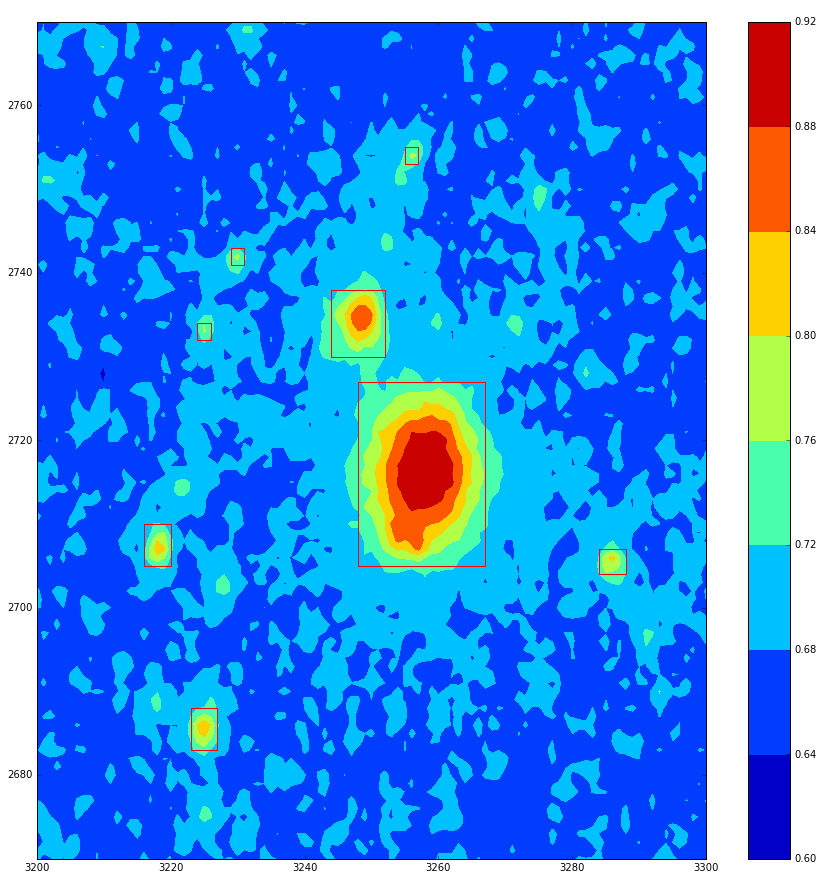}}
	\includegraphics[width=0.32\textwidth]{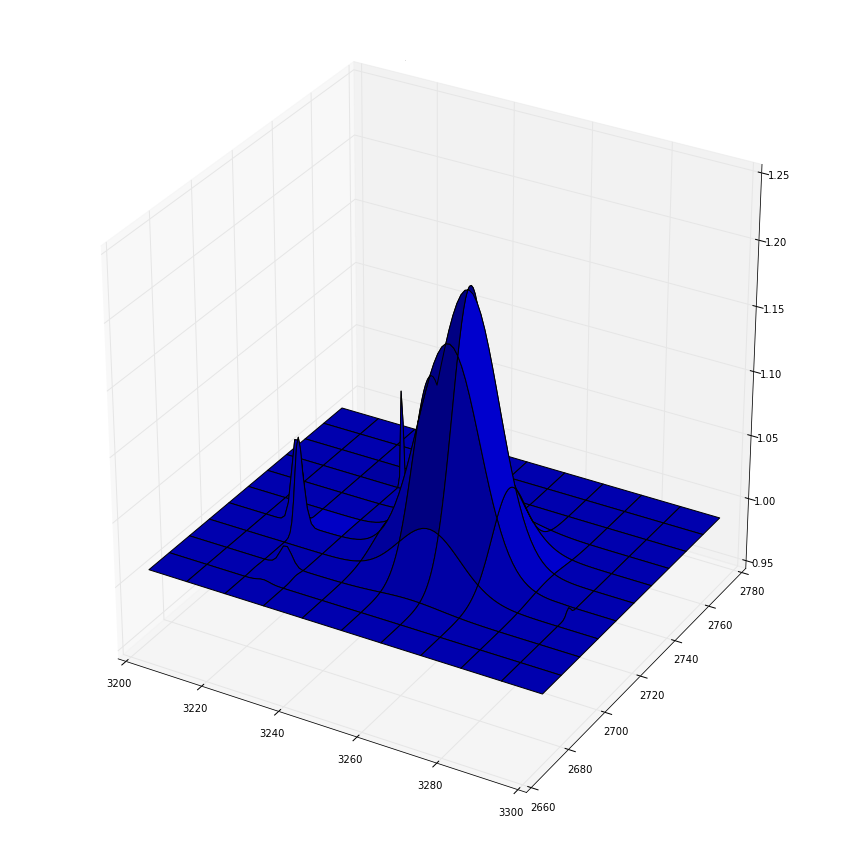}
	\includegraphics[width=0.32\textwidth]{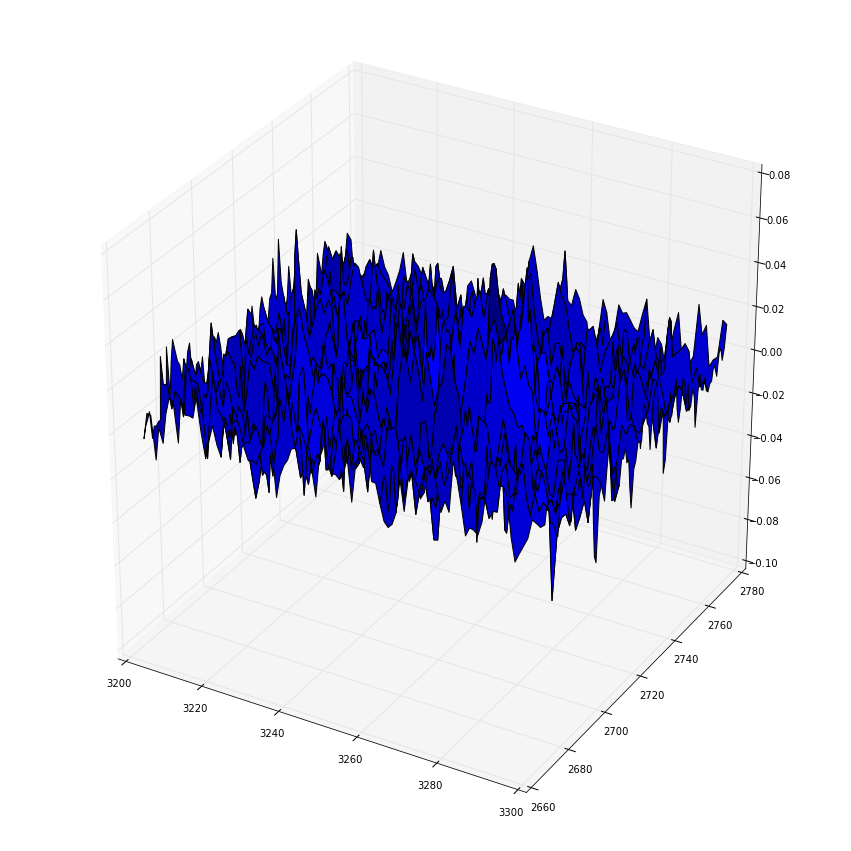}
	\caption{Adapted PSF fitting. Left: labeled stars in original digitized photographic plate. Center: model of the fitted function. Right: residuals after subtracting the fitted model. }
	\label{apf}
\end{figure*}

After background estimation, finding and labeling the stars, the data are exponentially
transformed, in order to use the built-in Gauss2D function within the model. In fig.\ \ref{apf} 
an example of a fit to a small section of one of the plates is shown.

The results are very promising.
Subtracting the model from the data we found an average of the residuals of 0.2\% with a standard deviation of 1.4\%, which is in good agreement with the standard deviation of the background of 1.1\%.

\section{Search for transient phenomema}

The Sonneberg plate archive furthermore might turn out to be a valuable database of transient phenomena of the 20th century, like novae, asteroids, etc.. In order
to develop an automated scheme for searching such events in the plate archive we started with a simple approach: a good quality plate of a series was chosen as a reference, which then was subtracted from all images of the series. To take into account fluctuations of the plate quality a high threshold was set to identify signals within the differences of the images. This method is quite similar to Pickering's positive--negative method \cite{Hoffleit1972}.

We analyzed the blue sensitive plates (labeled ''pg'') of the ''20h+40'' field, i.e. the center of the field is close to $\gamma$ Cygni. The images were cut in sections of 1\degr $\times$ 1\degr, to get rid of background drifts across the field. In a first step the centers of these sections were chosen from the position of known nova events during the observation campaign.

Several nova events were easily identified within the investigated series. As an example we present here our results of the Nova Cygni 1986 (V1819).This nova was discovered by M. Wakuda, Japan, 1986, August 4.7 as star of $m_V = 9.5$ mag \cite{iauc4242}. It was classified as a slow nova, the visible magnitude was  $m_V = 13$ mag on July 28.6 \cite{Duerbeck1987}, the nova reached its maximum of  $m_V = 8.5$ mag on August 9, 1986 \cite{Withney1989}.

\begin{figure*}[htp!]
	\centering
	\includegraphics[width=0.32\textwidth]{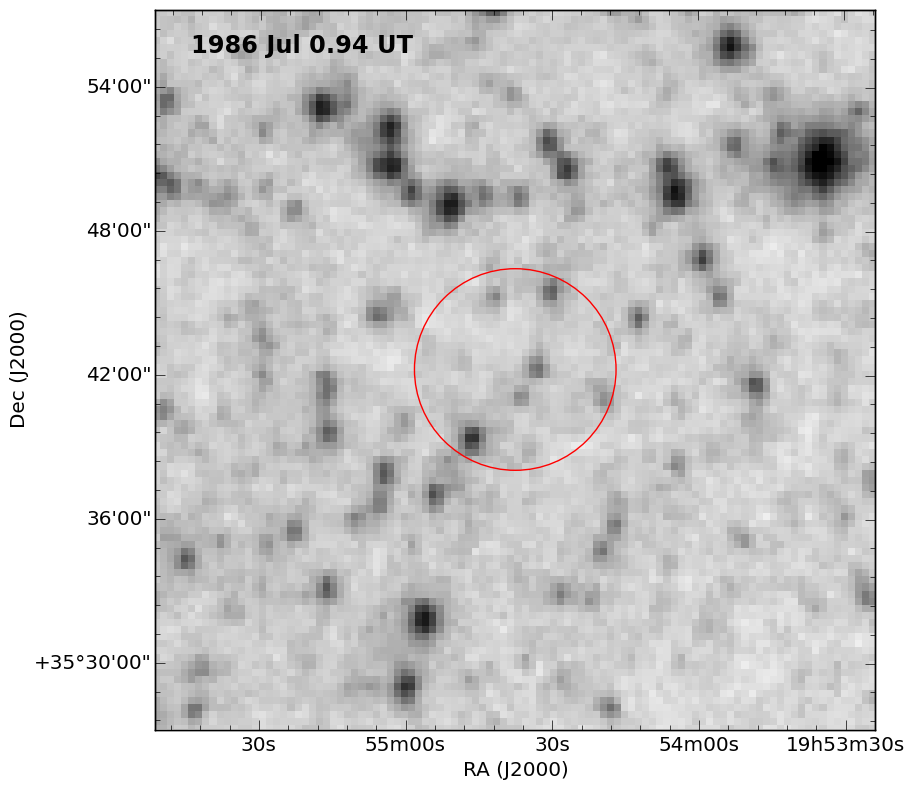}
	\includegraphics[width=0.32\textwidth]{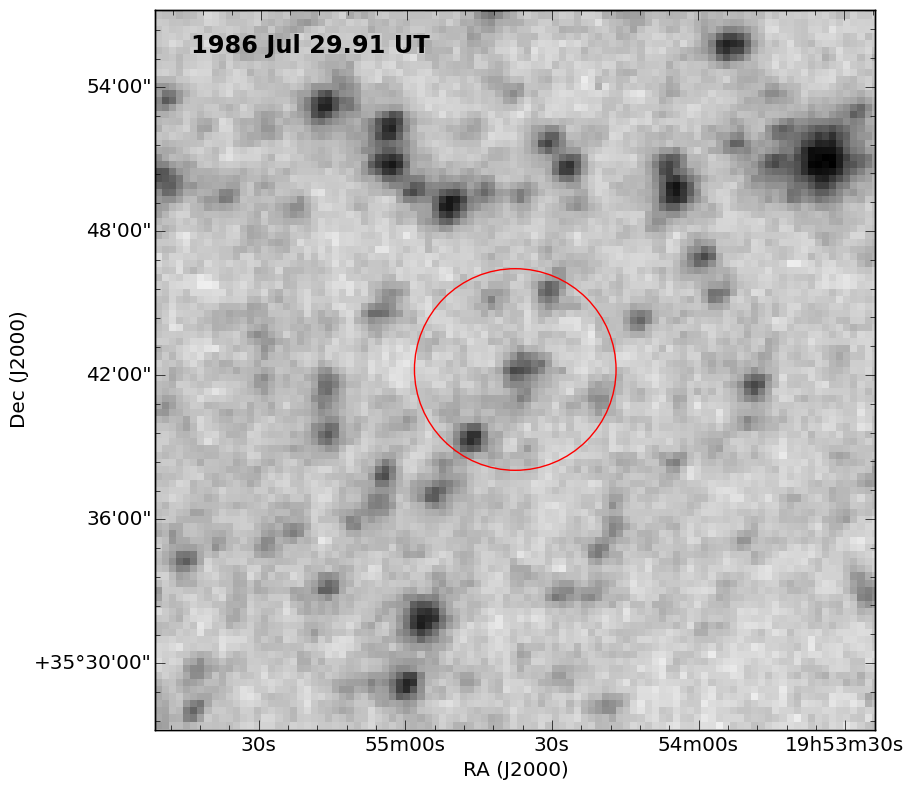}
	\includegraphics[width=0.32\textwidth]{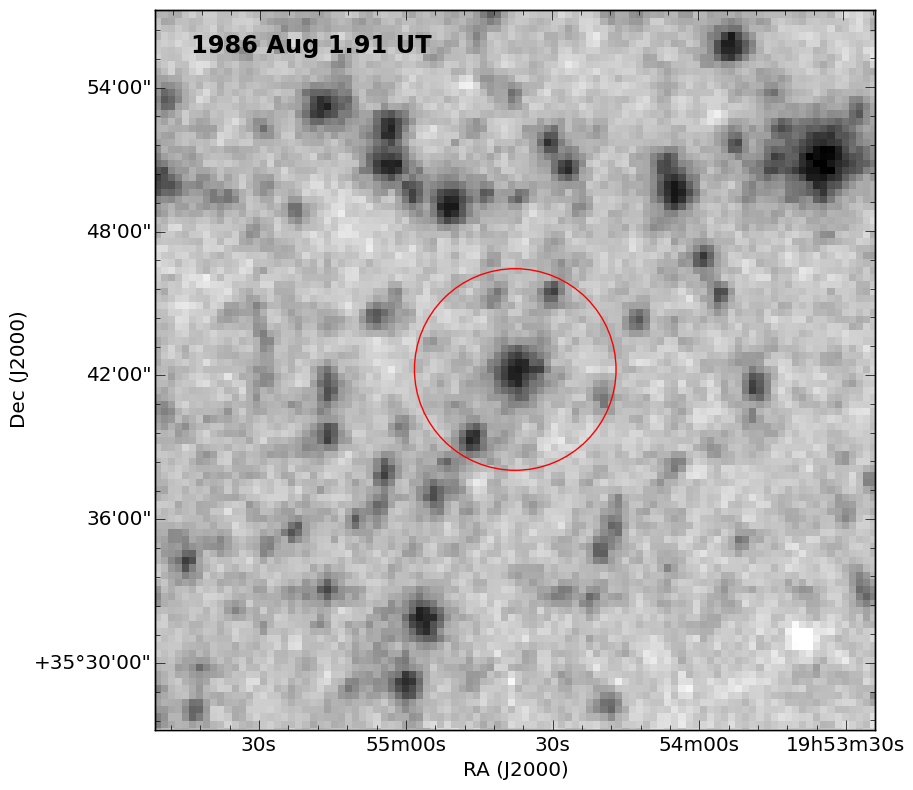}
	\caption{Nova Cygni 1986 (V1819) in the Sonneberg Skypatrol plates. Left: latest photo plate before the nova, July 1st. Center: the first appearance of the nova on July, 29th. Right: The nova's first maximum on Aug, 2nd.}
	\label{v1819}
\end{figure*}
We find a steep rise in brightness between July 29th and Aug 2nd, and in the blue sensitive series the brightness reached its maximum of about   $m_B = 9.5$ mag on Aug 2nd. This was already noted by I.I. Andronov et al., who manually analyzed this nova event in the same series of the Sonneberg plates \cite{Andronov1986}, and it proofs, that our procedure is capable of detecting nova events in the Sonneberg Skypatrol plates!

Within the analyzed sections of the series we additionally found several transient events on one or two following images only. These events clearly could be distinguished from scratches, but could not be identified with nova events. Further investigations are necessary to classify these events.

We are aware, that we might miss transient events with setting a high threshold. But the results so far are very promising. The next steps will be to normalize the images, to estimate the quality of the images e.g. from the statistics of the background and then to analyze the differences to the reference images while variing the threshold.

\end{document}